\documentclass[useAMS,usenatbib]{mn2e}
\usepackage{psfig}

\title[S5 0716+714]{Simultaneous MITSuME $g^{\prime}R_cI_c$ monitoring of S5 0716+714}
\author[C. S. Stalin et al. ]{C. S. Stalin$^{1}$\thanks{E-mail: stalin@iiap.res.in}, 
Koji. S. Kawabata$^{2}$, Makoto Uemura$^{2}$, Michitoshi Yoshida$^{3}$,
\newauthor 
Nobuyuki Kawai$^{4}$,  Kenshi Yanagisawa$^{3}$, Yasuhiro Shimizu$^{3}$,  
Daisuke Kuroda$^{3}$, \newauthor Shogo Nagayama$^{3}$, Hiroyuki Toda$^{3}$ \\
$^{1}$Indian Institute of Astrophysics, Koramangala, Bangalore 560\,034, India\\
$^{2}$Hiroshima Astrophysical Science Center, Hiroshima Univ., 1-3-1 Kagamiyama, Higashi-Hiroshima, Hiroshima 739-8526, Japan\\
$^{3}$Okayama Astrophysical Observatory, National Astronomical Observatory of
Japan, 037-5 Honjo, Asakuchi, Okayama 719-0232, Japan \\
$^{4}$Department of Physics, Tokyo Institute of Technology, 2-12-1 Ookayama, 
Meguro-ku, Tokyo 152-8551 \\
}

\begin{document}

\date{Accepted 1988 December 15. Received 1988 December 14; in original form 1988 October 11}

\pagerange{\pageref{firstpage}--\pageref{lastpage}} \pubyear{2008}

\maketitle

\label{firstpage}

\begin{abstract}
We present results of our intra-night
optical flux monitoring observations of S5 0716+714 done simultaneously 
in $g^{\prime}$, $R_C$ and $I_C$ filters. The observations were done using
Multicolor Imaging Telescopes for Survey and Monstrous Explosions (MITSuME)
instrument on the 50 cm telescope at the Okayama Astrophysical
Observatory over 30 nights
between 11 March 2008 and 8 May 2008. Of these
30 nights, 22 nights have continuous (without any break)
observations with duration ranging from
1 to 6 hours and hence were considered for intra-night optical variability
(INOV). The remaining 8 nights have continuous observations of
less than 1 hr and hence were considered only for long term optical variability
(LTOV).  In total we have 4888 datapoints which were simultaneous
in $g^{\prime}$, $R_C$ and $I_C$ filters. Of the 22 nights considered 
for INOV, the object showed flux variability  on 19 nights with the 
amplitude of variability 
in the $I_C$-band ranging from $\sim$4\% to $\sim$55\%. The duty cycle for INOV was 
thus found to be 83\%.
A good correlation between the light curves in all the three
bands was found. No time lag between different 
bands was noticed on most of the nights, 
except for 3 nights where the variation in $g^{\prime}$ was found to lead that
of the $I_C$ band by 0.3 to 1.5 hrs. On inter-night timescales, no lag 
was found between $g^{\prime}$ and $I_C$  bands.
On inter-night timescales as well as intra-night timescales on most of the
nights, the amplitude of variability was found to 
increase toward shorter wavelengths. The flux variations in the different bands
were not achromatic, with the blazar tending to become
bluer when brighter both on inter-night and intra-night
timescales; and this might be attributed to the larger amplitude
variation at shorter wavelengths. A clear periodic variation of 
3.3 hrs was found on 1 April 2008 and a hint for another possible periodic
variability of 4 hrs was found on 31 March 2008.
During our 30 days of observations over a 2 month period the source
has varied with an amplitude of variability as large as $\sim$80\%.

\end{abstract}


\section{Introduction}
Blazars form a sub-group of radio loud AGN showing extreme variability
at all wavelengths, high degrees of  linear polarization and strong
gamma ray emission. They include BL Lac objects as well as quasars
with flat radio spectra. They show variability over a wide range of 
timescales both within a night and over longer terms. Several models
have been proposed to explain their variability; the most
commonly accepted one is the shock-in-jet model (Marscher \& Gear 1985).
Alternative models that probably apply under specific circumstances involve 
interstellar scintillation (Rickett et al. 
2001), microlensing (Schneider \& Weiss 1987), accretion disk 
instabilities (Mangalam \& Wiita 1993; Chakrabarti \& Wiita 1993) and binary black holes (Sillanpaa 
et al. 1988). However, the mechanism responsible for variability
is not yet known conclusively.

Although optical variability on intra-night timescales is now a well
established phenomenon for blazars (Miller et al. 1989; Carini et al. 1992; Noble et al. 1997; 
Stalin et al. 2005;
Sagar et al.  2004), its relationship to long-term variability remains 
unclear. Clues to this relationship could possibly come from monitoring the optical spectrum for 
correlation with brightness (Vagnetti et al. 2003). This will also enable 
better discrimination 
among the various models proposed for flux variability in blazars.

Since the early times of BL Lac research, a correlation between spectral slope and source intensity has
been searched for (Gear
et al. 1986).  Several authors have studied the relationship of spectral 
changes to flux variations over the recent years (Takalo \& Sillanpaa 1989; 
D'Amicis et al. 2002; Vagnetti et al. 2003; 
Fiorucci et al. 2004; Foschini et al. 2006; Zheng et al. 2007).
D'Amicis et al. (2002)
reported that the spectra of all the eight BL Lacs in their sample
showed a bluer when brighter trend. Similar results were obtained
by Fiorucci et al. (2004) and Gu et al. (2006) also found that all
the BL Lacs in their sample tend to be bluer when brighter. However
some blazars are found to show anomalous spectral behaviour (Ramirez
et al. 2004). For example, PKS 0736+017 showed a tendency for its
spectrum to become redder when brighter both on inter-night and
intra-night timescales. Gu et al. (2006) found two of their
three FSQSR to be redder when they are brighter. Villata et al. (2006)
reported that the FSQSR 3C 454.3 generally had the redder when brighter
behaviours during the 2004$-$2005 outburst.  No  bluer when brighter trend was 
noticed either on intra-night or inter-night timescales for S5 0716+714 (Stalin
et al. 2006), whereas BL Lac showed a bluer when
brighter trend on intra-night intervals and a similar trend (although of less significance)
on inter-night timescales (Stalin et al. 2006; Papadakis et al. 2007).

It is generally considered that the optical spectral index
variability and the bluer when brighter trend are common for BL Lac 
objects (Gu et al. 2006; Papadakis et al. 2007).  
Nevertheless, it is currently not clear if this trend is universal in blazars
as some flat spectrum quasars show the opposite behaviour.
This bluer when brighter trend could be easily explained 
if the luminosity increase was due to the injection of fresh electrons with 
an energy distribution harder than that of previously partially cooled 
ones (Kirk et al. 1998; Mastichiadis \& Kirk 2002). A second explanation 
for that trend posits short term fluctuations of only the electron injection
spectral index (Bottcher \& Reimer 2004).

As blazars are variable in
timescales as short as minutes, the colour variations of blazars available in
the literature all of which are based on quasi-simultaneous multi-band 
monitoring have relatively
less utility than those obtained by simultaneous multiband
monitoring. Thus there is a need to have simultaneous monitoring observations of 
a sample of blazars to really resolve the bluer when brighter/redder when brighter
issue.  Here we report new observations
on the blazar S5 0716+714, which is one of the most studied
sources for variability across the electromagnetic spectrum (Wagner
et al. 1996; Kraus et al. 2003; Bach et al 2005; 
Bach et al. 2006). This blazar also has been observed repeatedly in various 
multifrequency campaigns (Wagner et al. 1990; Wagner et al. 1996; 
Tagliaferri et al.  2003; Raiteri et al. 2003) and
is known to be extremely variable ($\sim$hours to months at
radio and X-ray bands).  In the optical S5 0716+714 is identified 
as a BL Lac but with an unknown redshift. A redshift of z$>$ 0.3
was deduced using the limits on the surface brightness of the host
galaxy (Quirrenbach et al. 1991; Stickel et al. 1993; 
Sbarufatti et al. 2005). Nilsson et al. (2008) claim {\it z} = 0.31 $\pm$ 0.08
for S5 0716+714. 

The structure of the paper is as follows. Section 2 discusses the observations
and reductions. The analysis of the data is detailed in Section 3 and the
results are given in the final section.
\section{Observations and reductions}
Observations were carried out using the MITSuME 3 band ($g^{\prime}R_CI_C$) simultaneous imager
on the 50 cm telescope at the Okayama Astrophysical Observatory (Kotani et al. 2005). 
Observations were done over a total of 
30 nights with duration of observations ranging from 1 to 6 hours.

MITSuME has three dichroic mirrors to divide the incident beam
into three ones. It has been known that these dichroic mirrors
(inclined to the incident beam by 45 degrees) produces instrumental 
polarization of, at most, $1.2$\%, $2.0$\% and $6.2$\% at
$g'$, $R_{C}$ and $I_{C}$ bands, respectively.
This gives a photometric uncertainty up to $1.2p$\%, $2.0p$\%
and $6.2p$\%, respectively, where $p$ is the degree of the
intrinsic polarization of the object. However, the observed
polarization of S5~0716+714 during the observation period was
$\leq 0.15$ (Mahito Sasada et al., private communication) and thus the
estimated error is negligibly small, $\leq 0.003$ mag
even in $I_{C}$ band. 

\begin{table*}
\centering
 \begin{minipage}{175mm}
  \caption{Log of {\bf observations} of S5 0716+714. Here Npts is the number of data points and the number in parenthesis
in this column is the number of data points in 6 minute interval. Q$-$C and $\sigma_{Q-C}$ are the 
average differential magnitudes and their associated errors between the blazar and the comparison star. Similarly
C$-$K and $\sigma_{C-K}$ are the average differential magnitudes and errors respectively between the check star
and the comparison star}
  \begin{tabular}{@{}llllllllllllll@{}}
  \hline
Date        & Npts & \multicolumn{4}{c}{$g^{\prime}$} &  \multicolumn{4}{c}{$R_C$} &  \multicolumn{4}{c}{$I_C$}  \\ 
            &         & Q-C    & $\sigma_{Q-C}$ & C-K  & $\sigma_{C-K}$ & Q-C  & $\sigma_{Q-C}$ & C-K  & $\sigma_{C-K}$ & Q-C   & $\sigma_{Q-C}$ & C-K  & $\sigma_{C-K}$ \\ 
            &         & (mag)  & (mag)          & (mag)& (mag)          & (mag)& (mag)          & (mag)& (mag)          & (mag) & (mag)          & (mag)& (mag)          \\ \hline
11/03/2008  & 240(49) & 0.769 & 0.069 & 0.280 & 0.010 & 0.779 & 0.065 & 0.125 & 0.004 & 0.617 & 0.062 & 0.038 & 0.003  \\
12/03/2008  & 235(49) & 0.470 & 0.070 & 0.278 & 0.009 & 0.501 & 0.052 & 0.123 & 0.005 & 0.350 & 0.044 & 0.035 & 0.004  \\
14/03/2008  & 269(47) & 0.475 & 0.020 & 0.304 & 0.011 & 0.492 & 0.017 & 0.137 & 0.008 & 0.332 & 0.018 & 0.044 & 0.005  \\
16/03/2008  & 211(41) & 0.452 & 0.045 & 0.279 & 0.037 & 0.474 & 0.033 & 0.131 & 0.025 & 0.316 & 0.027 & 0.041 & 0.007  \\
17/03/2008  & 448(63) & 0.669 & 0.042 & 0.281 & 0.012 & 0.645 & 0.033 & 0.122 & 0.005 & 0.467 & 0.034 & 0.035 & 0.004  \\
20/03/2008  &  78(11) & 0.839 & 0.030 & 0.303 & 0.006 & 0.841 & 0.024 & 0.125 & 0.003 & 0.673 & 0.024 & 0.038 & 0.004  \\
21/03/2008  & 356(52) & 0.615 & 0.053 & 0.304 & 0.013 & 0.637 & 0.045 & 0.133 & 0.004 & 0.480 & 0.044 & 0.045 & 0.004  \\
22/03/2008  & 326(52) & 0.589 & 0.054 & 0.284 & 0.018 & 0.639 & 0.051 & 0.123 & 0.007 & 0.497 & 0.043 & 0.032 & 0.003  \\
24/03/2008  & 122(18) & 0.447 & 0.023 & 0.294 & 0.008 & 0.468 & 0.020 & 0.125 & 0.004 & 0.315 & 0.017 & 0.037 & 0.003  \\
25/03/2008* &  31(6)  & 0.584 & 0.019 & 0.301 & 0.009 & 0.586 & 0.013 & 0.132 & 0.003 & 0.420 & 0.006 & 0.041 & 0.002  \\
26/03/2008  & 246(36) & 0.620 & 0.038 & 0.296 & 0.018 & 0.650 & 0.029 & 0.127 & 0.005 & 0.512 & 0.088 & 0.039 & 0.003  \\
28/03/2008  & 90(14)  & 0.465 & 0.018 & 0.296 & 0.007 & 0.486 & 0.015 & 0.126 & 0.003 & 0.328 & 0.015 & 0.040 & 0.002  \\
31/03/2008  & 280(43) & 0.776 & 0.029 & 0.295 & 0.007 & 0.762 & 0.027 & 0.131 & 0.006 & 0.572 & 0.024 & 0.044 & 0.003  \\
01/04/2008  & 292(45) & 0.940 & 0.023 & 0.288 & 0.007 & 0.928 & 0.019 & 0.125 & 0.004 & 0.741 & 0.018 & 0.038 & 0.003  \\
02/04/2008* &  32(8)  & 0.738 & 0.177 & 0.268 & 0.119 & 0.834 & 0.143 & 0.151 & 0.060 & 0.574 & 0.025 & 0.041 & 0.020  \\
03/04/2008  & 118(19) & 0.792 & 0.033 & 0.289 & 0.016 & 0.792 & 0.031 & 0.127 & 0.004 & 0.610 & 0.030 & 0.036 & 0.004  \\
04/04/2008  & 121(19) & 0.601 & 0.017 & 0.287 & 0.008 & 0.610 & 0.013 & 0.127 & 0.005 & 0.436 & 0.012 & 0.035 & 0.003  \\
06/04/2008  & 123(18) & 0.877 & 0.024 & 0.285 & 0.010 & 0.855 & 0.025 & 0.126 & 0.005 & 0.662 & 0.025 & 0.035 & 0.004  \\
08/04/2008* &  42(7)  & 0.610 & 0.029 & 0.289 & 0.005 & 0.615 & 0.030 & 0.129 & 0.004 & 0.442 & 0.028 & 0.038 & 0.004  \\
11/04/2008* &  40(8)  & 0.552 & 0.020 & 0.287 & 0.011 & 0.528 & 0.014 & 0.121 & 0.010 & 0.350 & 0.014 & 0.040 & 0.005  \\
12/04/2008* &  32(7)  & 0.556 & 0.013 & 0.299 & 0.011 & 0.507 & 0.027 & 0.125 & 0.010 & 0.355 & 0.046 & 0.038 & 0.003  \\
25/04/2008  & 242(35  & 0.450 & 0.018 & 0.285 & 0.007 & 0.445 & 0.016 & 0.120 & 0.005 & 0.276 & 0.017 & 0.033 & 0.003  \\
26/04/2008  & 187(27) & 0.465 & 0.012 & 0.289 & 0.005 & 0.467 & 0.008 & 0.122 & 0.004 & 0.303 & 0.009 & 0.035 & 0.004  \\
27/04/2008* &  44(6)  & 0.169 & 0.011 & 0.304 & 0.017 & 0.174 & 0.008 & 0.123 & 0.008 & 0.016 & 0.011 & 0.049 & 0.008  \\
28/04/2008* &  79(12) & 0.400 & 0.040 & 0.291 & 0.012 & 0.412 & 0.010 & 0.141 & 0.026 & 0.227 & 0.053 & 0.046 & 0.003  \\
29/04/2008  & 204(32) & 0.369 & 0.012 & 0.294 & 0.012 & 0.383 & 0.011 & 0.129 & 0.007 & 0.219 & 0.012 & 0.044 & 0.004  \\
02/05/2008  &  66(16) & 0.213 & 0.058 & 0.296 & 0.025 & 0.237 & 0.049 & 0.137 & 0.009 & 0.072 & 0.050 & 0.053 & 0.009  \\
05/05/2008  & 136(24) & 0.578 & 0.009 & 0.320 & 0.010 & 0.562 & 0.006 & 0.151 & 0.010 & 0.365 & 0.009 & 0.067 & 0.008  \\
07/05/2008  & 131(22) & 0.555 & 0.029 & 0.295 & 0.017 & 0.543 & 0.024 & 0.138 & 0.009 & 0.360 & 0.022 & 0.058 & 0.009  \\
08/05/2008* &  67(12) & 0.690 & 0.034 & 0.307 & 0.023 & 0.651 & 0.023 & 0.131 & 0.013 & 0.475 & 0.016 & 0.053 & 0.009  \\ 
  \hline
\end{tabular}
\end{minipage}
\end{table*}

\begin{table*}
\centering
 \begin{minipage}{175mm}
  \caption{Results of flux monitoring of S5 0716+714}
  \begin{tabular}{@{}lllllllllllll@{}}
  \hline
Date         & T     & C       & Status & \multicolumn{3}{c}{$g^{\prime}$} & \multicolumn{3}{c}{$R_C$} & \multicolumn{3}{c}{$I_C$} \\ 
             &       &         &     & $D_{min}$ & $D_{max}$ & Amp.  & $D_{min}$ & $D_{max}$  & Amp. & $D_{min}$ & $D_{max}$  & Amp. \\ 
             & (hrs) &         &     & (mag)     & (mag)     & (\%)  & (mag)     & (mag)      & (\%) & (mag)     & (mag)      & (\%) \\  \hline
 11/03/2008  & 4.9 & 20.667  &  V  & 0.665 & 0.855 & 18.95 & 0.677 &  0.857 & 17.99 &  0.523 &  0.695 & 17.19  \\
 12/03/2008  & 5.1 & 11.000  &  V  & 0.365 & 0.615 & 24.97 & 0.426 &  0.614 & 18.79 &  0.282 & 0.454 & 17.19 \\
 14/03/2008  & 5.2 &  3.600  &  V  & 0.435 & 0.509 &  7.23 & 0.465 &  0.528 &  6.20 &  0.299 & 0.376 &  7.67 \\
 16/03/2008  & 5.0 &  3.857  &  V  & 0.249 & 0.534 & 28.02 & 0.380 & 0.536  & 15.19 &  0.263 & 0.415 & 15.17 \\
 17/03/2008  & 6.3 &  8.500  &  V  & 0.600 & 0.756 & 15.51 & 0.587 & 0.706  & 11.88 &  0.417 & 0.543 & 12.59 \\
 20/03/2008  & 1.1 &  6.000  &  V  & 0.791 & 0.876 &  8.46 & 0.809 & 0.867  &  5.78 &  0.634 & 0.702 &  6.78 \\
 21/03/2008  & 5.2 & 11.000  &  V  & 0.556 & 0.783 & 22.63 & 0.585 & 0.786  & 20.09 &  0.431 & 0.621 & 18.99 \\
 22/03/2008  & 5.2 & 14.333  &  V  & 0.493 & 0.685 & 19.03 & 0.556 & 0.725  & 16.87 &  0.428 & 0.571 & 14.29 \\
 24/03/2008  & 1.8 &  5.667  &  V  & 0.410 & 0.488 &  7.72 & 0.440 & 0.500  &  5.97 &  0.289 & 0.345 &  5.58 \\
 26/03/2008  & 3.7 & 29.333  &  V  & 0.572 & 0.735 & 16.10 & 0.580 & 0.699  & 11.88 &  0.458 & 1.011 & 55.30 \\
 28/03/2008  & 1.4 &  7.500  &  V  & 0.436 & 0.488 &  5.10 & 0.465 & 0.508  &  4.28 &  0.305 & 0.348 &  4.29 \\
 31/03/2008  & 4.4 &  8.000  &  V  & 0.726 & 0.832 & 10.55 & 0.714 & 0.822  &  10.77&  0.531 & 0.613 &  8.19 \\
 01/04/2008  & 4.5 &  6.000  &  V  & 0.892 & 0.974 &  8.14 & 0.894 & 0.953  &  5.87 &  0.710 & 0.773 &  6.29 \\
 03/04/2008  & 2.9 &  7.500  &  V  & 0.736 & 0.839 & 10.05 & 0.745 & 0.838  &  9.28 &  0.564 & 0.658 &  9.38 \\
 04/04/2008  & 2.9 &  4.000  &  V  & 0.569 & 0.629 &  5.89 & 0.582 & 0.630  &  4.75 &  0.413 & 0.462 &  4.88 \\
 06/04/2008  & 2.9 &  6.250  &  V  & 0.838 & 0.914 &  7.47 & 0.820 & 0.894  &  7.37 &  0.630 & 0.701 &  7.08 \\
 25/04/2008  & 3.5 &  5.667  &  V  & 0.420 & 0.490 &  6.93 & 0.424 & 0.477  &  5.25 &  0.253 & 0.310 &  5.68 \\
 26/04/2008  & 2.7 &  2.250  & NV  & 0.430 & 0.483 &  5.25 & 0.449 & 0.482  &  3.25 &  0.286 & 0.320 & 3.35       \\
 29/04/2008  & 3.2 &  3.000  &  V  & 0.344 & 0.389 &  4.17 & 0.363 & 0.406  &  4.18 &  0.196 & 0.245 &  4.87 \\
 02/05/2008  & 4.3 &  5.556  &  V  & 0.096 & 0.285 & 18.57 & 0.136 & 0.301  & 16.45 & -0.019 & 0.131 & 14.95 \\
 05/05/2008  & 2.4 &  1.125  & NV  & 0.555 & 0.596 & 3.85  & 0.552 & 0.574  & 1.69  & 0.350  & 0.384 & 3.21  \\
 07/05/2008  & 2.2 &  2.444  & NV  & 0.499 & 0.595 & 9.29  & 0.502 & 0.577  & 7.39  & 0.321  & 0.403 & 8.10   \\
  \hline
\end{tabular}
\end{minipage}
\end{table*}

The observational errors are estimated from the rms differential
magnitudes between the two comparison stars

\begin{equation}
\sigma = \sqrt{\frac{(m_i - \bar{m})^2}{N-1}}
\end{equation}

\noindent where $m_i$ = $m_c - m_k$ is the differential magnitude of 
the comparison
star and the check star, $\bar{m}$ is the differential magnitude
averaged over the entire data set, and N is the number of observations
on a given night. For each night the typical rms error ranges between 
0.003  mag in $I_C$ to 0.01 mag in $g^{\prime}$ band. The log of observations
along with the average differential magnitudes between the quasar and 
the comparison star, comparison star and check star and their associated
errors in $g^{\prime}R_CI_C$ bands are given in Table 1. An asterisk against
dates in Table 1 denotes the nights for which the duration of continuous
observation is less than 1 hour and are not considered for intra-night
optical variability  analysis.  Those nights are only considered for
long term variability analysis. 

\begin{figure*}
\psfig{file=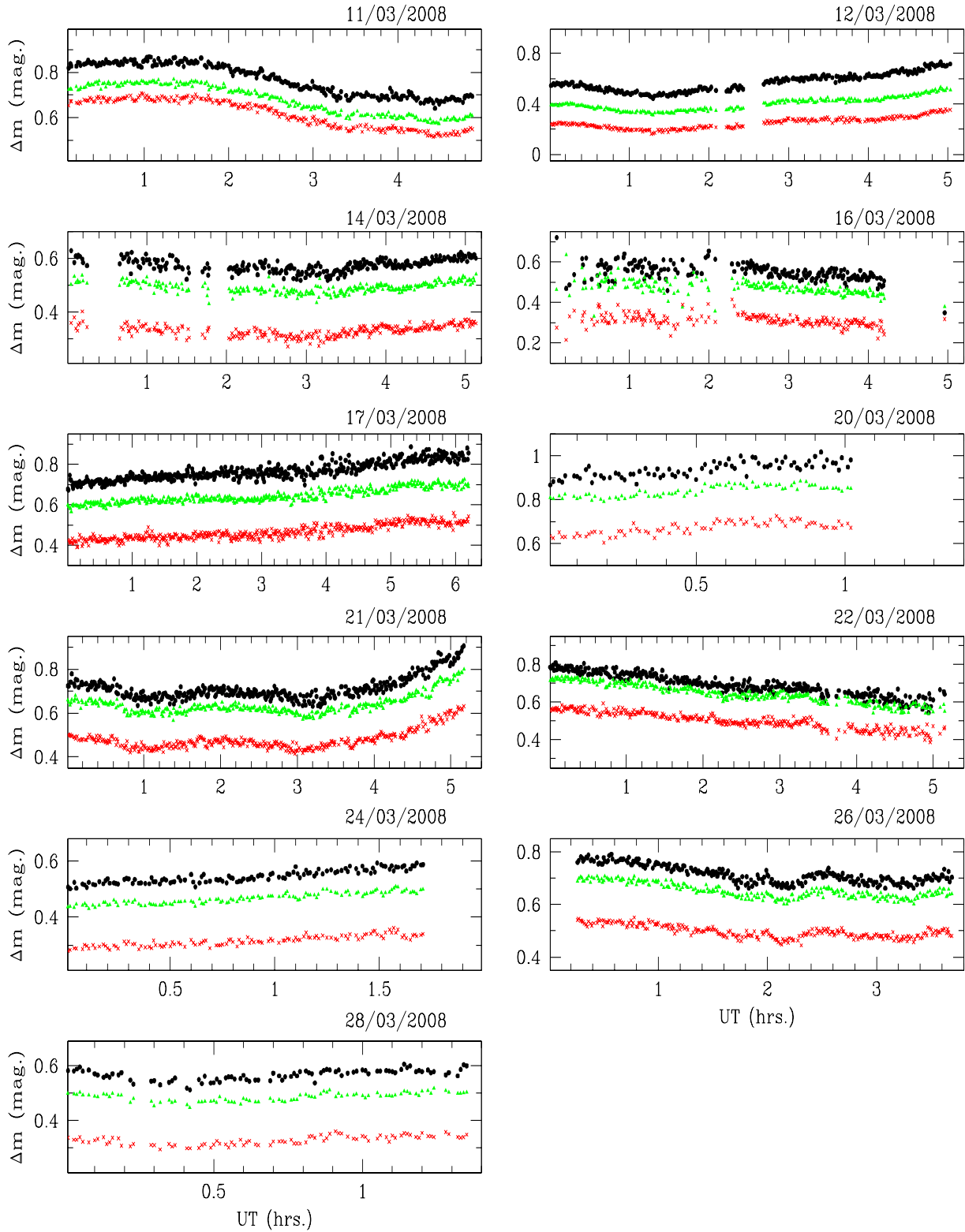,width=18cm,height=23cm}
\caption{\sf Differential light curves (DLCs) of S5 0716+714 in 
$g^{\prime}R_CI_C$ filters. {\bf The DLCs in $g^{\prime}$, $R_C$ and $I_C$ 
bands are shifted differently on each night and then shown here
for clarity.} The dates of 
observations are indicated on each panel. Here filled squares are the 
$g^{\prime}$-band observations {\bf (top)}, filled triangles are 
$R_C$-band observations {\bf (middle)}
and crosses are $I_C$-band observations {\bf (bottom)}.}
\end{figure*}

\begin{figure*}
\psfig{file=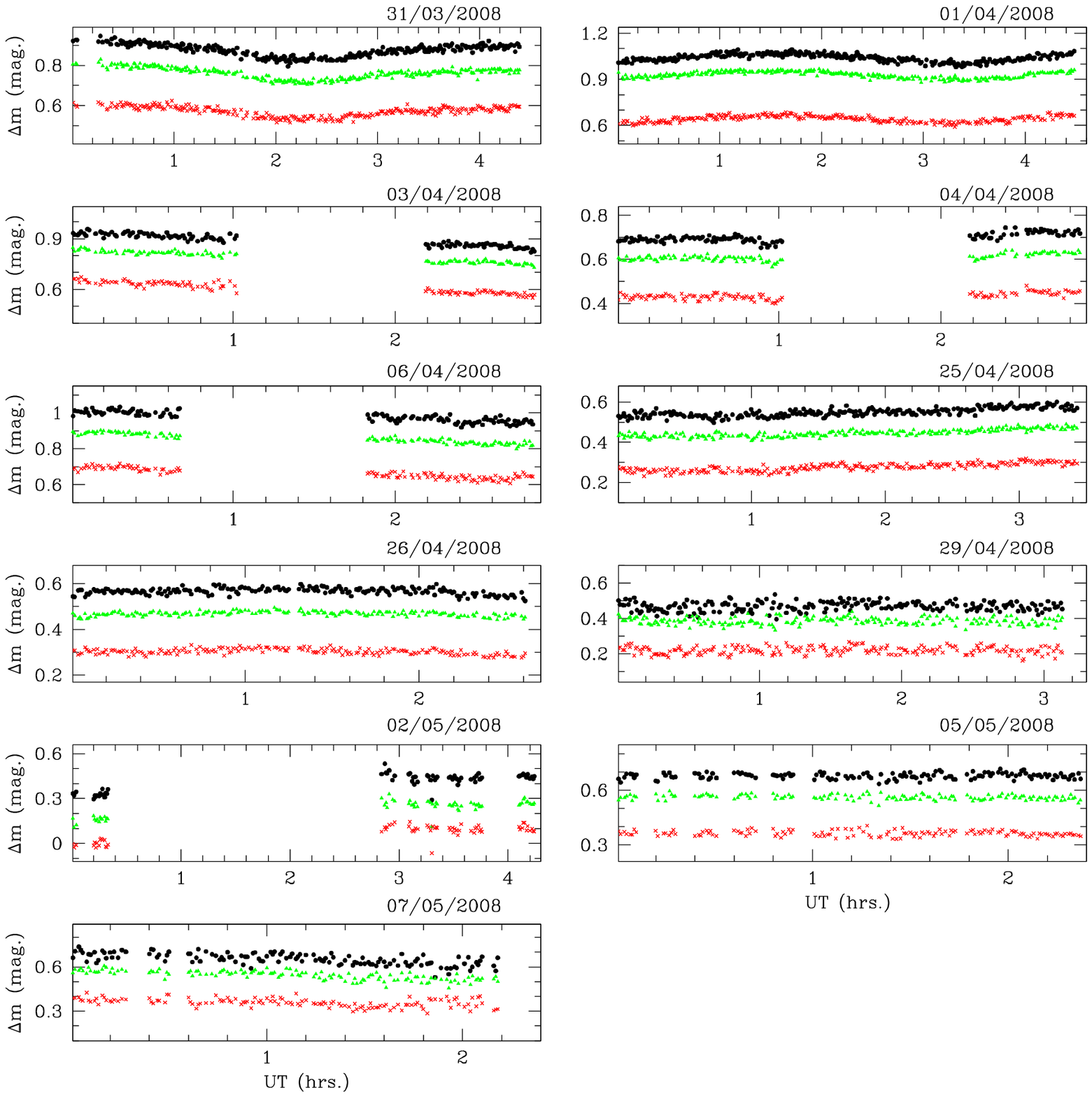,width=18cm,height=23cm}
\caption{\sf Differential light curves of S5 0716+714 in $g^{\prime}R_CI_C$ filters. {\bf Details to the figure are the same as in Fig. 1.}}
\end{figure*}

\section{Analysis}
\subsection{Intra-night optical variability (INOV)}
A total of 30 nights of observations were carried out on the source.
Of these, data on 8 nights were noisy and also have continuous duration
of observations less than 1 hour. These 8 nights were hence
used only for long term optical variability (LTOV) analysis. Thus for INOV analysis we consider 
only 22 nights. To check for variability, we first constructed
two differential light curves (DLCs), one between the blazar and
a comparison star (C) and the other between the comparison star (C)  and a 
check star (K).  Comparison and check stars are chosen such that they are 
non-variable on each particular night. 
To say if S5 0716+714 has shown INOV in a given night, we have
employed the statistical criterion of Jang \& Miller (1997). This
is based on a parameter C defined as C = $\frac{\sigma_{QSO-C}}{\sigma_{C-K}}$. 
The source is classified as variable on any given 
night if C $>$ 2.567 and non-variable otherwise. This corresponds to 
$>$ 99\% confidence level in variability.
{\bf To test for the variability of the source, we have used only the $I_C$ band
data as it has the largest S/N compared to the data in 
$g^{\prime}$ and $R_C$ bands. On each night, the typical rms error 
in $I_C$-band was 0.003 mag. This is much lower compared to the
corresponding rms error values of 0.007 and 0.01 in $R_C$ and $g^{\prime}$
bands respectively}.
The values of $C$ estimated from the $I_C$ band, the 
nightly variability status,  duration of observations
and nightly variability amplitudes are given in Table 2. 
From Table 2 it is clear that of the 22 nights 
considered for INOV, the object is clearly variable on 19 nights. 
The DLCs of the object on all the 22 nights are given in Figures 1 and 2.
{\bf However, we also estimated the C parameter for $g^{\prime}$ and $R_C$ bands
as well. The object passed the variability criteria in all the three bands
for a total of 11 nights. On 5 nights it was variable on both
$R_C$ and $I_C$ bands and on 3 nights it was variable when only the $I_C$ band
data was considered}.


\subsubsection{Variabiltiy amplitude} 
We define the amplitude of variability following Romero et al. (1999)

\begin{equation}
\psi = \sqrt{(D_{max} - D_{min})^2 - 2\sigma^2}
\end{equation}
where $D_{min}$ and $D_{max}$ refer to the minimum and maximum in the 
differential light curve of the object relative to a comparison star
present on the same CCD frame
and $\sigma^2$ is the standard deviation in the differential light curve of
the comparison and check stars as given in Eq. 1. As the errors are subtracted
from the total measured variability, this expression for $\psi$ gives a fairer
estimate of the true amplitude of variability in the source. The values of
$\psi$ for each nights of observations in the three bands are given in Table 2.

\subsubsection{Duty cycle of variability}
The duty cycle for S5 0716+714 is calculated as (Romero et al. 1999) 

\begin{equation}
DC = 100 \frac{\sum_{i=1}^n N_i(1/\Delta t_i)}{\sum_{i=1}^N(1/\Delta t_i)} \%
\end{equation}
where $\Delta t_i$ = $\Delta t_{i,obs} (1+z)^{-1}$ is the duration
of the observation of the source corrected for the cosmological
redshift of the $i_{th}$ night. $N_i$ equals 0 if the object was
non-variable and 1 if it was variable during $\Delta t_i$. On the 22
nights having duration of observations between 1 and 6 hrs considered
for INOV, the object was variable on 19 nights. This leads to an estimation
of the duty cycle of variability of $\sim$83\%. This is similar to the 
duty cycle of INOV shown by blazars (Stalin et al. 2004).

\subsubsection{Cross correlations}
We  have  used the Discrete Correlation function (DCF) method of
Edelson \& Krolik (1988) to check for any time lag between 
$g^{\prime}$ and $I_C$ bands. Here we first calculated a set of unbinned DCFs
defined as

\begin{equation}
UDCF_{ij} = \frac{(a_i - \bar{a})(b_j - \bar{b})}{\sigma_a*\sigma_b}
\end{equation}

where $a_i$ and $b_j$ are the observed differential magnitudes in the
two different filters and $\bar{a}$, $\bar{b}$, $\sigma_a$ and $\sigma_b$ are respectively
the mean and standard deviation of the DLCs in the respective filters. 
DCF($\tau$) is obtained
by binning the results in $\tau$. Averaging over
M pairs for 
which $\tau - \delta\tau/2 \le \delta t_{ij} < \tau + \delta \tau/2$ gives 

\begin{equation}
DCF(\tau) = \sum_{i=1}^{M}UDCF_{ij}/M
\end{equation}

The errors in the DCF are estimated using

\begin{equation}
\sigma_{DCF}(\tau) = \frac{1}{(M-1)}\sum_{i=1}^N[UDCF_{ij} - DCF(\tau)]^2
\end{equation}

The position of the maximum in the DCF is estimated using 
the centroid ($\tau_c$) of the DCF, given by

\begin{equation}
\tau_c = \frac{\sum_{i} \tau_i DCF_i}{\sum_{i} DCF_i}
\end{equation}

This centroid is estimated for points which are within 80\% of the peak
value of the DCF. {\bf Some examples of} the computed DCF between $g^{\prime}$ and $I_C$ band 
are shown in Fig. 3 . Of the 19 nights the object showed INOV, there is 
a clear indication of a time lag between $g^{\prime}$ and $I_C$ bands 
with durations of 0.3 to 1.6 hrs on 3 nights. {\bf However, on other nights
the non-existence of a time lag might be because of
correlating a poor quality $g^{\prime}$ lightcurve with a relatively better
quality $I_C$ band lightcurve}. The DCF peak and centroid values
are given in Table 3.

\begin{figure}
\psfig{file=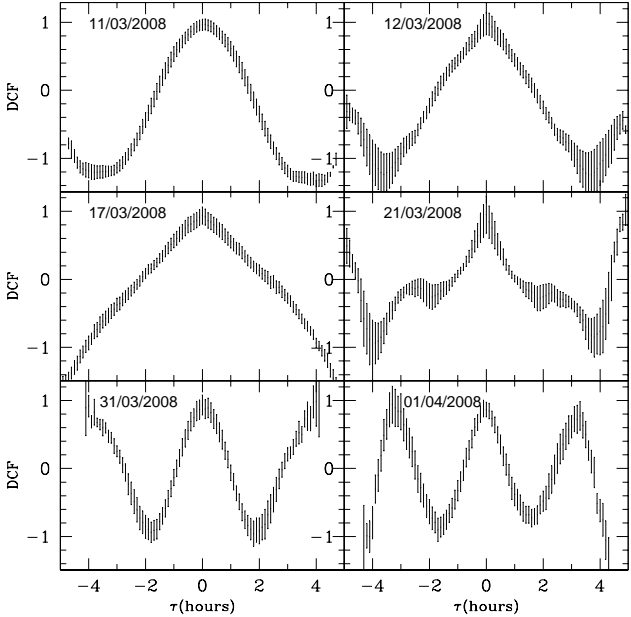,width=9cm,height=9cm}
\caption{\sf {\bf Some examples of} Discrete Correlation Function (DCF) 
between the $g^{\prime}$ and
$I_C$ filters for S5 0716+714 {\bf on} the nights when intra-night optical 
variabiltiy was observed. The dates of observations are indicated
on each panel.}
\end{figure}

\begin{figure}
\psfig{file=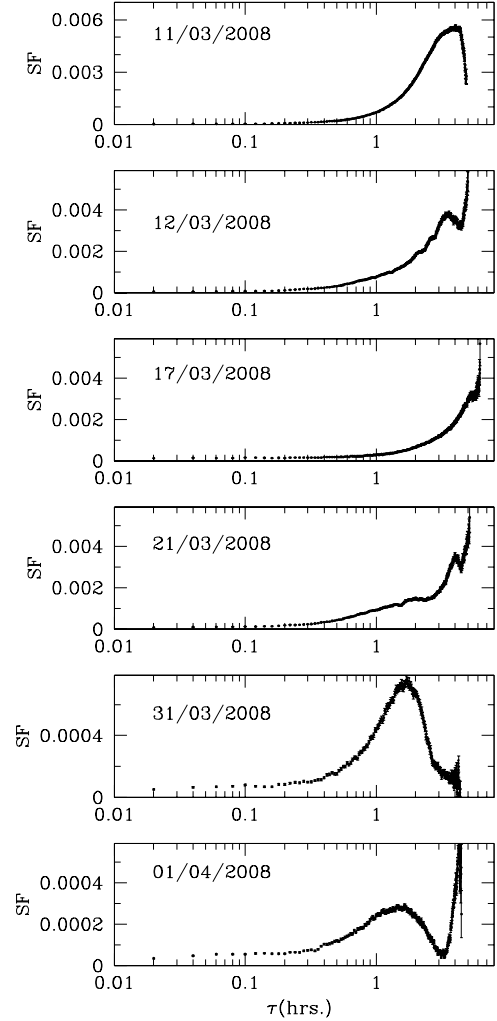,width=9cm}
\caption{\sf {\bf Some examples of} structure function {\bf plots} 
of S5 0716+714 on the nights
when INOV was observed. The dates of observations are given on each panel.}
\end{figure}

\subsubsection{Variability timescale and periodicity}
The presence of periodic or quasi-periodic variations in 
the lightcurves of blazars may provide evidence for
accretion disk based models like accretion disk pulsation 
(Chakrabarti \& Wiita 1993) and orbiting hot spot on 
the accretion disk (Mangalam \& Wiita 1993). {\bf In jet based
models too, periodic variations can arise from jet 
precession and jet rotation (Reiger 2004; Camenzind \& 
Krockenberger 1992; Gopal-Krishna \& Wiita 1992)}. Some reports on the 
presence of periodic or quasi-periodic variations in blazars are
available in literature. Gupta et al. (2009) found good
evidence for optical periodic variations in S5 0716+714 ranging from
$\sim$25 to $\sim$73 minutes. Few other blazars also have shown
evidence for periodic variations in their optical lightcurves.
In OJ 287, on intra-night timescales, 23 min and 32 min periodicity
were reported by Carrasco et al. (1985) and Carini et al. (1992)
respectively. On inter-night timescales, a quasi-periodicity of
0.7 days seemed to be present in PKS 2155$-$304 (Urry et al. 1993).
Hints for hour like timescale periodic variations were reported for 
the blazars 0851+202, 0846+513 and 1216+010 by Stalin et al. (2004). To look for
possible periodicities in the new observations on S5 0716+714, we
have used the first order structure function. The first order
structure function is also used to find the variability time scale
of the source.  This is defined as (see Simonetti et al. 1985) 

\begin{equation}
D_X^1(\tau) = \frac{1}{N(\tau)} \sum_{i=1}^N [X(i + \tau) + X(i)]^2
\end{equation}

\noindent where $\tau$ is the time lag, $N(\tau)$ = $\sum w(i)w(i + \tau)$, the 
weighting function $w(i)$ is equal to 1 if a  measurement exists for the $i_{th}$ interval and
0 otherwise. Each point in the SF is associated with an error defined as

\begin{equation}
\sigma(\tau)^2 = \frac{8\sigma^2_{\delta X}}{N(\tau)} D^1_X(\tau)
\end{equation}

\noindent where, $\sigma^2_{\delta X}$ is the measured noise variance.
To estimate the SF, the data set was first transformed into uniform
intervals sampled every 6 min.  A typical time scale in the light 
curve (i.e., the time between
a maximum and a minimum or vice versa) shows up as a local
maximum in the SF. In the case of a monotonically increasing SF, the
source possesses no typical time-scale smaller than the total
duration of observations. A minimum in the SF indicates the presence of 
possible periods in the light curve. {\bf For SF analysis 
also we have used the $I_C$ band data as it has the lowest 
errors of the three bands}. Nightly variability timescales
ranging from 0.1 to 5.3 hrs were found for the source.
From SF analysis, on the 19 nights the object
has shown variability, hints of quasi-periods were found on 9 nights. 
Of these 9 nights, clear evidence for periodic variation with 3.3 hrs
 was found on 1 April 2008. Also there is evidence of a possible periodic
variation with a period of 4.0 hrs on 31 March 2008. This is cleary
seen in the lightcurves shown in Fig.2 and is further supported by
the Discrete Correlation Function analysis of the $I_C$ band 
data with itself (autocorrelation). The presence of 
strong peaks in autocorrelation apart from the one near zero indicate a 
periodicity. The autocorrelation was also similar to the DCF  plots 
shown in Fig.3. The results of the SF are given in Table 3 and {\bf few
examples of} the SF 
plots are shown in Fig. 4.

\subsubsection{Colour Variations}
From the differential instrumental magnitudes of the blazar relative to 
the comparison star, standard magnitudes of the blazar were obtained
considering the standard magnitudes of $g^{\prime}$ = 12.45 mag., 
$R_C$ = 12.08 mag. and $I_C$ = 11.76 mag. for the comparison star. To check for
colour evolution, the $g^{\prime}-I_C$ colours were computed and plotted against
the $g^{\prime}$-band magnitudes. The colour magnitude diagrams for all
the 19 nights when the object showed INOV are shown in Fig. 5. 
Also shown in Fig. 5 are the unweighted linear least squares fit
to the data. Results of this linear regression analysis are given in Table 4. 
Clear evidence for a bluer when brighter trend was found on most of the nights. 

\subsection{Inter-night variability}
The DLCs showing the inter-night variability are shown in Fig. 6.
The source {\bf was} found to vary upto 0.8 mag during the period of 
observations between 11 March 2008 and 8 May 2008.
On inter-night time scales too, the amplitude of variability {\bf was} 
found to increase toward shorter wavelengths as can be
seen in Table 5.  The object has shown correlated variability in 
all the three bands in inter-night timescales as well.  
The DCF between $g^{\prime}$ and
$I_C$ bands on inter-night timescales is shown in Fig. 7. We
found no lag between the $g^{\prime}$ and
$I_C$ bands. In Fig. 8 is shown the colour ($g^{\prime}-I_C$) magnitude ($I_c$)
diagram of S5 0716+714 on inter-night timescales along with an unweighted linear least
squares fit to the data.  A  bluer when brighter trend was found. 

\begin{table}
\centering
 \begin{minipage}{80mm}
  \caption{Results of structure function and 
           discrete correlation function analysis}
  \begin{tabular}{@{}lrcrr@{}}
  \hline
Date       & $\tau$     & Period  & DCF peak & DCF Centroid  \\ 
           & (hrs.)      & (hrs.) & (hrs.)   &  (hrs.)  \\ \hline
11/03/2008 & 4.0     &           & 0.10 &  0.09\\
12/03/2008 & 2.1,2.6 & 4.3       & 0.00 &  0.05\\
14/03/2007 & 2.0     & 3.9       & 0.00 &  0.00\\
16/03/2008 & ----    & ----      & 0.03 &  0.18\\
17/03/2008 & 5.3     &           & 0.00 & -0.05\\
20/03/2008 & 0.6     & 0.7,0.9   & 0.00 & -0.09\\
21/03/2008 & 1.4,1.8 & 1.6,2.1   & 0.00 &  0.00\\
22/03/2008 & 3.6     &           & 0.00 &  0.05\\
24/03/2008 & 1.3     & 1.4       & 0.00 & -0.05 \\
26/03/2008 & $>$ 3.5 &           & 0.00 &  0.00 \\
28/03/2008 & 0.8     &           & 0.00 &  0.28 \\
31/03/2008 & 1.7     & 4.0       & 0.00 &  0.00\\
01/04/2008 & 1.5     & 3.3       &-0.10 &  0.00 \\
03/04/2008 & 0.1     & 0.9       & 0.00 &  0.00 \\
04/04/2008 &         &           & 0.90 &  0.85\\
06/04/2008 & 2.4     &           &-0.20 & -0.10 \\
25/04/2008 & 2.9     &           & 0.10 & -0.04 \\
29/04/2008 &         &           & 1.60 &  1.60 \\
02/05/2008 & 2.7     & 2.9       & 0.00 &  0.00 \\
  \hline
\end{tabular}
\end{minipage}
\end{table}

\begin{table}
\centering
 \begin{minipage}{140mm}
  \caption{Correlation between the $g^{\prime}$$-$$I_C$ and $g^{\prime}$}
  \begin{tabular}{@{}llll@{}}
  \hline
Date       & Slope            & intercept         & R     \\ \hline
11/03/2008 & 0.12 $\pm$ 0.01  &  -0.79 $\pm$ 0.16 & 0.54  \\
12/03/2008 & 0.39 $\pm$ 0.01  &  -4.22 $\pm$ 0.13 & 0.93  \\
14/03/2007 & 0.43 $\pm$ 0.04  &  -4.66 $\pm$ 0.55 & 0.50  \\
16/03/2008 & 0.84 $\pm$ 0.05  & -10.01 $\pm$ 0.59 & 0.74  \\
17/03/2008 & 0.38 $\pm$ 0.02  &  -4.10 $\pm$ 0.31 & 0.60  \\
20/03/2008 & 0.49 $\pm$ 0.07  &  -5.66 $\pm$ 0.94 & 0.64  \\
21/03/2008 & 0.24 $\pm$ 0.01  &  -2.36 $\pm$ 0.22 & 0.59  \\
22/03/2008 & 0.32 $\pm$ 0.02  &  -3.39 $\pm$ 0.26 & 0.68  \\
24/03/2008 & 0.33 $\pm$ 0.04  &  -3.39 $\pm$ 0.49 & 0.60  \\
26/03/2008 & 0.39 $\pm$ 0.02  &  -4.24 $\pm$ 0.27 & 0.73  \\
28/03/2008 & 0.40 $\pm$ 0.06  &  -4.27 $\pm$ 0.82 & 0.56  \\
31/03/2008 & 0.30 $\pm$ 0.03  &  -3.01 $\pm$ 0.34 & 0.56  \\
01/04/2008 & 0.40 $\pm$ 0.04  &  -4.50 $\pm$ 0.47 & 0.54  \\
03/04/2008 & 0.17 $\pm$ 0.04  &  -1.42 $\pm$ 0.58 & 0.35  \\
04/04/2008 & 0.53 $\pm$ 0.05  &  -6.08 $\pm$ 0.70 & 0.64  \\
06/04/2008 & 0.21 $\pm$ 0.06  &  -1.93 $\pm$ 0.76 & 0.30  \\
25/04/2008 & 0.42 $\pm$ 0.05  &  -4.59 $\pm$ 0.58 & 0.49  \\
29/04/2008 & 0.92 $\pm$ 0.06  & -10.93 $\pm$ 0.75 & 0.80  \\
02/05/2008 & 0.19 $\pm$ 0.05  &  -1.56 $\pm$ 0.59 & 0.45  \\
  \hline
\end{tabular}
\end{minipage}
\end{table}

\begin{figure*}
\psfig{file=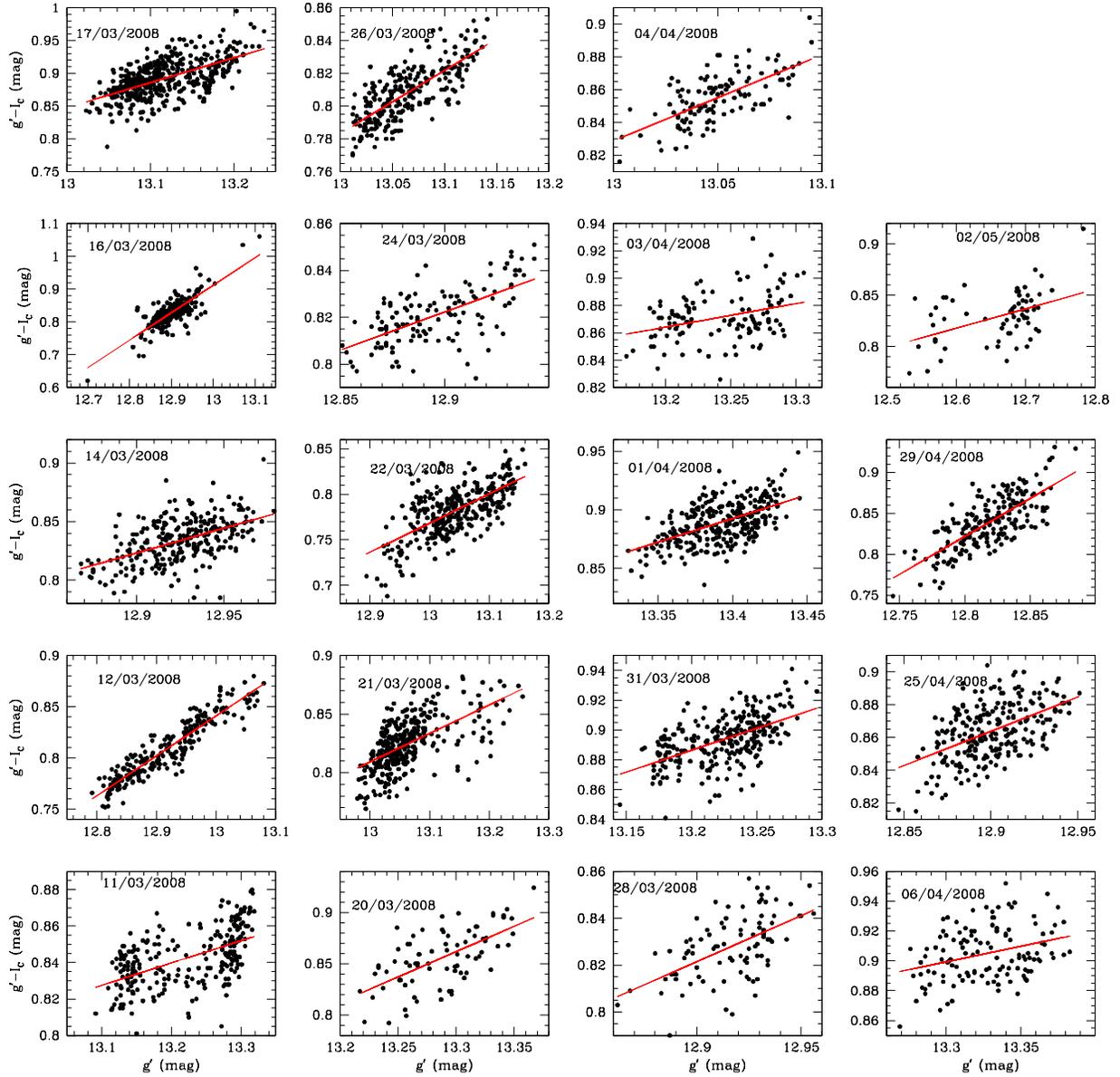,width=18cm,height=18cm}
\caption{\sf Colour magnitude diagram of S5 0716+714 for the nights where
intra-night optical variability (INOV) was found. The dates are indicated
on each panel. {\bf The solid line is the linear
least squares fit to the data}.}
\end{figure*}

\begin{table}
\centering
 \begin{minipage}{140mm}
  \caption{Inter-night optical variability statistics}
  \begin{tabular}{@{}lllll@{}}
  \hline
Filter  & $D_{min}$  & $D_{max}$  &  $\sigma_{C-K}$  & Amplitude  \\ 
        & (mag)      & (mag)      &  (mag)           & (\%)       \\ \hline
$g^{\prime}$       & 0.145 $\pm$ 0.055    &  0.940 $\pm$ 0.023   & 0.011     &  79.5 \\
$R_C$       & 0.174 $\pm$ 0.008    &  0.928 $\pm$ 0.019   & 0.008     &  75.4 \\
$I_C$       & 0.016 $\pm$ 0.011    &  0.741 $\pm$ 0.018   & 0.008     &  72.5 \\
  \hline
\end{tabular}
\end{minipage}
\end{table}

\section{Summary}
The presence or absence of a bluer when brighter trend in blazars on inter-night
and intra-night timescales can provide interesting clues to the origin of blazar
activity from hour like to much longer timescale. Such a study on the 
spectral variability in blazars also will help to constrain various models
proposed for blazar activity. This relationship between the optical spectral
variability and brightness variations in blazars have been investigated by
many authors (Speziali \& Natali 1998; Papadakis et al. 2003; Raiteri et al. 2003;
Villata et al. 2004; Wu et a. 2005; Stalin et al. 2006). The results of
such studies are contrary to each other. To address this issue, we have presented
here high temporal resolution, simultaneous $g^{\prime}R_CI_C$ band photometric
monitoring observations of the blazar S5 0716+714 on 30 nights between
11 March 2008 and 8 May 2008.  The results of our observations are summarized
as follows

\begin{enumerate}
\item The object was active during our whole monitoring period
and showed variability both on intra-night and inter-night timescales.
During individual nights, the amplitude of variability ranges
from 4\% to 55\%. 
 On inter-night timescale, the source has {\bf shown}
a variability as large as 80\% during the whole monitoring period
\item Of the 22 nights considered for INOV, the object showed variability
on 19 nights with an estimated duty cycle of variability of 83\%. 
The amplitide of {\bf variability} was found to be larger toward
shorter wavelengths, both on inter-night and intra-night timescales
\item On the nights the object showed INOV, the timescale of variability
was found to be between {\bf 0.1} hr and 5.3 hrs. Also on 9 nights evidence
for quasi-periods were found with periods ranging from 0.9 to 4.3 hrs.
Clear evidence for periodic variations with a period of 3.3 hrs was found
on 1 April 2008. A second possible periodic variation with period of
4.0 hrs was found on 31 March 2008.
\item No evidence for time lag was found between $g^{\prime}$ and $I_C$ 
bands on most of the nights, except for three nights, when the $g^{\prime}$
band was found to lead the $I_C$  band with durations from 0.3 to 1.6 hrs. 
However, on inter-night timescales, no time lag was found between
$g^{\prime}$ and $I_C$ bands.
\item The object showed clear colour variation, in the sense
the object became bluer when brighter on both intra-night
and inter-night timescales.
\end{enumerate}
{\bf Several models have been proposed to explain the flux variability in blazars.
They are broadly grouped into two categories namely intrinsic and extrinsic.
Models invoking extrinsic mechanisms as the cause of variability include
interstellar scintillation (Rickett et al. 2001) and gravitational
microlensing (Schneider \& Weiss 1987; Gopal-Krishna \& Subramanian 1991).
Interstellar scintillation can be active at low radio freqencies as it is 
highly frequency dependent. The optical INOV seen in S5 0716+714 cannot thus
be caused by interstellar scintillation. Gravitational microlensing results
in symmetric lightcurves and is also an achromatic process. The clear 
bluer when brighter chromatism seen in S5 0716+714 both on inter-night and
intra-night timescales is thus not due to microlensing. These observations
as well as the close correlations between the optical and radio bands seen in 
S5 0716+714 (Quirrenbach et al. 1991) argues strongly against an extrinsic
origin of variability. The two other major classes of models for intrinsic
origin of AGN variability are those involving accrection disk 
instabilities (Mangalam \& Wiita 1993) and those involving shocks in 
relativistic jets (Marscher \& Gear 1985). The shock-in-jet model is the most
commonly used model to explain variability in blazars where relativistic jets
are present. In this model time lag between various wavelength bands are
expected. In our observations no evidence of a lag between $g^{\prime}$ and
$I_C$ bands was found on most of the nights. This might be due to the poor
quality of the $g^{\prime}$ band data compared to the $I_C$ band for
correlation analysis.  Investigations on flux variability on blazar sources,
have revealed evidence of periodicity in different timescales in few cases
(Gupta et al. 2009; Carrasco et al. 1985; Carini et al. 1992; Stalin et al.
2005, Wu et al. 2005). The detection of quasi-periodicity
or periodicity on intra-night timescales can be most explained by accretion
disk based models (Mangalam \& Wiita; Chakrabarti \& Wiita 1993; Espailat
et al. 2008). In the context of shock-in-jet model too, periodicity 
in blazar lightcurves can be of geometrical origin namely orbital motion
in a binary black-hole system, jet precession and jet rotation (Reiger 2004; 
Camenzind \& Krockenberger 1992; Gopal-Krishna \& Wiita 1992) and their
variations are achromatic (Wu et al. 2005). In our observations on the 
nights when quasi-periodicities were found and on the two nights when
clear sinusoidal variations were found, the variations were highly chromatic. 
The presence of spectral variations thus imply that the observed variations
are not caused by geometric effects. 
Our observations  on the variability in S5 0716+714, the bluer when 
brighter trend and the increase of the amplitude of variability towards
shorter wavelenghts appear to be more consistent in terms of 
shocks propagating in the relativistic jet of S5 0716+714.
Recently Dai et al. (2009) noted that among blazars, BL Lacs show a bluer
when brighter trend whereas, FSQSRs show a redder when brighter trend. 
This might indicate the existence of different physical conditions in these
two subclasses of blazars. Further simultaneous observations of a matched 
sample of BL Lacs and FSQSRs are needed to fully resolve the question of the 
ubiquity of the  bluer when brighter trend in blazars}.

\begin{figure}
\psfig{file=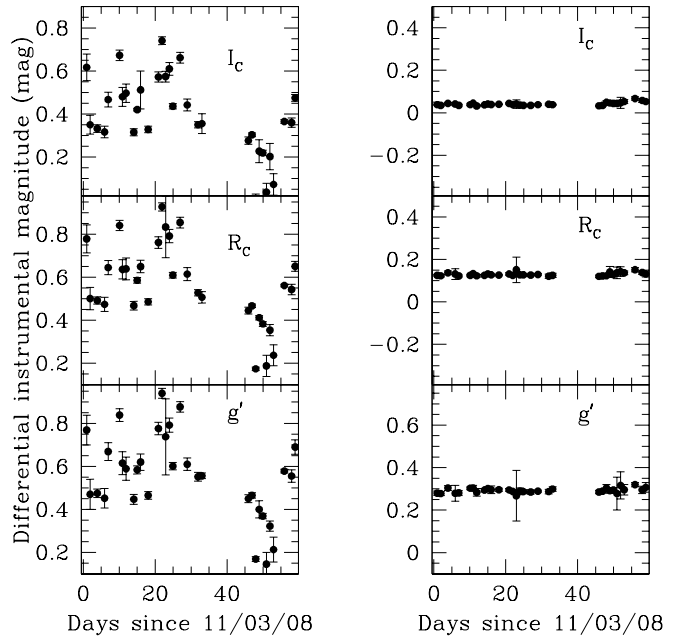,width=9cm,height=9cm}
\caption{\sf Inter-night optical variability of S5 0716+714 during
the period 11 March 2008 to 08  April 2008. Left: Differential Light
Curves in $g^{\prime}R_cI_c$ filters. Right: DLCs between the 
comparison star and the check star in $g^{\prime}R_cI_c$ filters filters.}
\end{figure}

\begin{figure}
\psfig{file=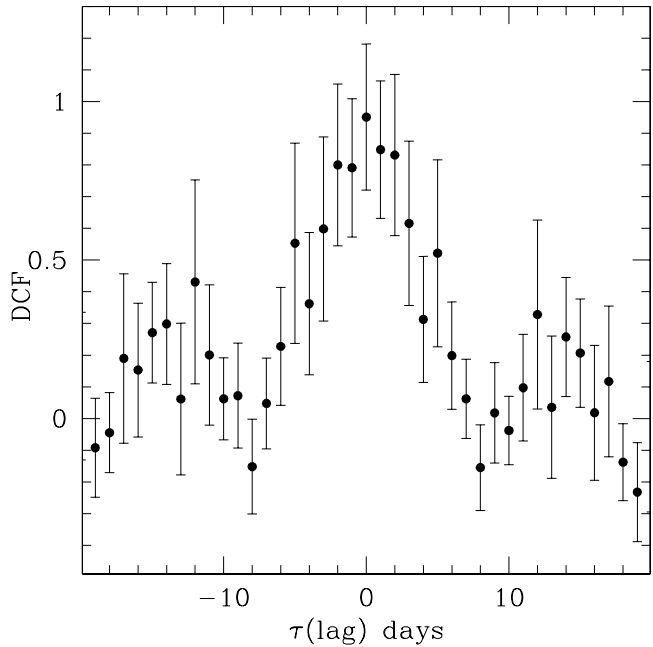,width=9cm,height=9cm}
\caption{\sf Discrete Correlation Function (DCF) between $g^{\prime}$ and
$I_c$ in S5 0716+714 on inter-night timescales}
\end{figure}

\begin{figure}
\psfig{file=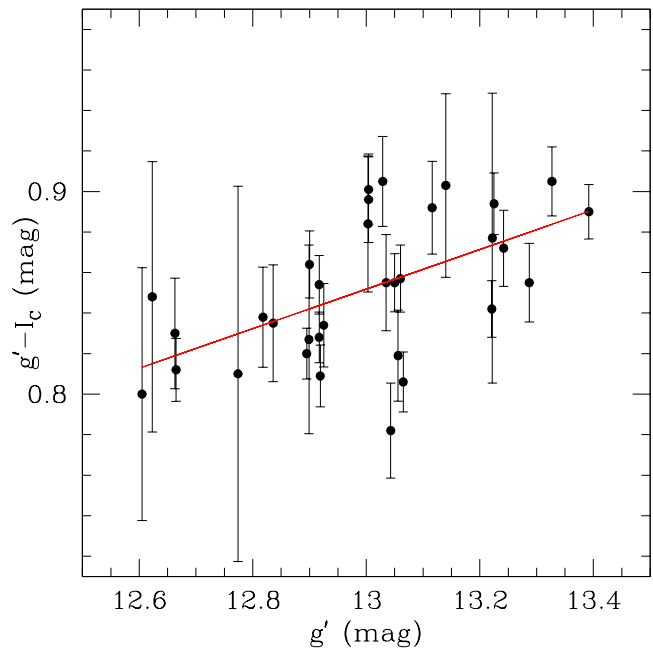,width=9cm,height=9cm}
\caption{\sf Colour-magnitude diagram on inter-night timescales}
\end{figure}

\section*{Acknowledgments}
We thank Prof. Paul J. Wiita for comments on the manuscript. We also thank
the anonymous referee for his/her comments that improved the paper.
This research has made use of the NASA/IPAC Extragalactic Database (NED) which 
is operated by the Jet Propulsion Laboratory, California Institute of 
Technology, under contract with NASA.

\end{document}